\documentclass[12pt]{iopart}

\usepackage{iopams}  
\usepackage{hyperref}
\usepackage{booktabs}
\hypersetup{colorlinks,linkcolor={blue},citecolor={blue},urlcolor={blue}} 
\usepackage{graphicx}
\begin{document}

\title[Author guidelines for IOP Publishing journals in  \LaTeXe]{Inverse Potentials for all $\ell$ channels of Neutron-Proton Scattering using Reference Potential Approach}

\author{Anil Khachi$^1$, Lalit Kumar$^1$, Ayushi Awasthi$^1$ and O. S. K. S. Sastri$^{1,*}$}

\address{Department of Physics and Astronomical Sciences\\ Central University of Himachal Pradesh\\ Dharamshala, 176215, Himachal Pradesh, Bharat(India)}
\ead{sastri.osks@gmail.com}
\vspace{10pt}
\begin{indented}
\item[]March 2023
\end{indented}

\begin{abstract}
Reference potential approach (RPA) is successful in obtaining inverse potentials for weakly bound diatomic molecules using Morse function. In this work, our goal is to construct inverse potentials for all available $\ell$-channels of np-scattering using RPA. The Riccati-type phase equations for various $\ell$-channels are solved using 5th order Runge-Kutta method to obtain scattering phase shifts (SPS) in tandem with an optimization procedure to minimize mean squared error (MSE). Interaction potentials for a total of 18 states have been constructed using only three parameter Morse interaction model. The obtained MSE is $<1\%$ for $^1S_0$, $^3P_1$ and $^3D_1$ channels and $<2\%$ for $^1P_1$ channel and $<0.1\%$ for rest of the 14 channels. The obtained total scattering cross-sections at various lab energies are found to be matching well with experimental ones. Our complete study of np-scattering for all $\ell$-channels using RPA using Morse function as zeroth reference, is being undertaken for the first time.
\end{abstract}
\section{Introduction}
The neutron-proton (\textit{n-p}) interaction has been first modeled by Yukawa \cite{Yukawa}. This was followed by various single and multi-particle exchange models and QCD based models as detailed in these reviews \cite{Naghdi, Machleidt}. Currently, the Nijmegen \cite{Nagels}, Argonne v18 \cite{Wiringa}, CD-Bonn \cite{Machleidt} and Granada\cite{Perez} potentials are the ones which give rise to best quantitative results for explaining the experimental scattering phase shifts. Unfortunately, all these potentials have different mathematical representations originating from completely varied physical considerations. Yet all of them lead to correct validation of experimental data. The search for a simple theoretical potential that could model the nucleon-nucleon interactions is still eluding the physicists. Interestingly, many simple phenomenological forms such as Square well, Malfiet-Tjon \cite{Malfliet}, Hulthen \cite{PRC}, have also been utilised for studying the deuteron. Recently,  molecular potentials such as Manning-Rosen \cite{Khirali}, Morse \cite{AnilChitkara} and Deng-Fan\cite{bhoi} have been proposed. 

Typically, phase wave analysis (PWA) technique based on modeling interaction potential, relies on solving time independent Schrodinger equation (TISE) to obtain scattering phase shifts (SPS) from which differential and total scattering cross-sections (SCS) are obtained and matched with experimental ones. While R-matrix\cite{Rmatrix}, S-matrix\cite{Smatrix}, complex scaling method (CSM)\cite{CSM}, etc rely on wavefunction to obtain SPS, phase function method (PFM) method\cite{PFM} directly utilises the model potential in phase equation to solve SPS directly. Zhaba \cite{Zhaba} utilised PFM method to study nucleon-nucleon scattering, and Laha et al.\cite{PRC}studied nucleon-nucleus and nucleus-nucleus interaction \cite{PRC} and obtained reasonably good results. 

%They mainly used molecular potentials like double Hulthen and Manning- Rosen phenomenological potentials as models of nuclear potentials. 

%On other hand PFM has been exploited by Zhaba and have directly realistic potentials like Reid93 and Argonne-v18 in it for studying np scattering. We instead in this paper are using well known Morse potential with PFM. It is more or less well established now that the strong force is a result of quark interactions within hadrons and mesons, and the nucleons themselves experience an attractive force which is of secondary nature akin to Vander Wall interaction within two neutral atoms in a molecule. Based on this premise and understanding of the characteristics of \textit{np} interaction which has a repulsive core at short inter-nucleon distance ($< 0.5 fm$), followed by an attractive nature between $0.5 - 2$ fm and an exponentially decaying tail as the nucleons go far apart, we have considered in our recent work \cite{Sastri}, the most successful molecular Morse potential to describe phenomenologically the \textit{np} interaction.\\

Alternatively, inverse potentials resulting directly from experimental observations by J-matrix method \cite{Zaitsev} and Marchenko equation \cite{Selg} have also found some success in understanding the interaction involved between the nucleons. Recently, an analytical procedure to solve this Marchenko equation has been achieved by modeling the interaction using a Morse curve \cite{Morse}, which belongs to a class of shape invariant potentials \cite{Morales}, called as reference potential approach (RPA). Selg\cite{Selg} points out that \textit{``The implementation of the methods of the inverse scattering theory is not at all a trivial task. On the contrary, this is a complex and computationally very demanding multi-step procedure that has to be performed with utmost accuracy"}. We have constructed inverse potentials for S-waves of np \cite{Sastri}, nD\cite{Shikha}, S, P and D-states of $n-\alpha$ \cite{Kumar L} and $\alpha-\alpha$ \cite{Khachi A} scattering by numerically solving the phase equation choosing Morse function as zeroth reference. The goal of this paper is to extend this reference potential approach (RPA) to constructing inverse potentials for all 18 $\ell$-channels of np-interaction by considering recent data for SPS from Arriola et.al., of Granada group \cite{Perez} for lab energies up to 350 MeV.

\section{Methodology:}
The Morse function is given by
\begin{equation}
V_{Morse}(r) = V_0\left(e^{-2(r-r_m)/a_m)}-2e^{-(r-r_m)/a_m}\right)
\label{eq1}
\end{equation}
The PWA approach to understanding two body scattering focuses on obtaining phase shift for various orbital angular momentum, $\delta_{\ell}$, between incoming and outgoing wave due to the interaction between projectile and target nucleons. It is important to note that an attractive potential would tend to pull the wave backward and hence results in a positive SPS, $\delta_\ell > 0$ and a repulsive potential would push it out which makes $\delta_{\ell} < 0 $. Typically, one solves radial TISE to obtain wavefunction from which SPS are deduced. One of the many advantages of Morse potential is that it's TISE can be analytically solved for, $\ell=0$, S-states for both bound and scattering states \cite{Matsumoto}.
\subsection{Analytical solution for Scattering State of Morse potential:}
Schr$\ddot{o}$dinger wave equation, for a spinless particle with energy ($E$) and orbital angular momentum ($\ell$) undergoing scattering from another particle with interaction potential V(r), is given by
\begin{equation}
\frac{\hbar^2}{2\mu} \bigg[\frac{d^2}{dr^2}+\big(k^2-\ell(\ell+1)/r^2\big)\bigg]u_{\ell}(k,r) = V(r)u_{\ell}(k,r)
\label{Scheq}
\end{equation}
Where $k=\sqrt{E/(\hbar^2/2\mu)}$. 
%For \textit{np} system, center of mass energy $E_{c.m.}$ is related to laboratory energy $E_{\ell ab}$  by following relation for non-relativistic kinematics 
%\begin{equation}
%E_{c.m.}=\frac{M_p}{M_p+M_n} E_{\ell ab.}
%\end{equation}
The analytical solution of TISE for Morse potential \cite{Morse} with $\ell = 0 $, is given by
\begin{equation}
E_v = -\frac{\hbar^2}{2\mu a_m^2}(\lambda -v-1/2)^2~~~~v = 0, 1, 2, \ldots
\label{eigenvals}
\end{equation}
where
\begin{equation}
\lambda = \sqrt{\frac{2\mu V_0 {a_m}^2}{\hbar^2}}~~{and}~~\frac{\hbar^2}{2\mu} = 41.47~MeV fm^{2}
\end{equation}
is called well-depth parameter and is dependent only on $V_0$ and $a_m$.
\subsubsection{SPS for $^1S_0$ singlet state:}
\noindent In case of singlet $^1S_0$ unbound state ($E > 0$), with $\ell = 0$, the analytical formula for 
\textit{SPS} due to Morse potential is obtained \cite{darewych1967morse, Matsumoto} as

\begin{equation}
\delta_0^{ana} = -k r_m - \epsilon(\gamma + log 2\lambda)+\sum_{n=1}^\infty \left(\frac{\epsilon}{n}- tan^{-1}\frac{2\epsilon}{n} + tan^{-1}\frac{\epsilon}{v-\lambda-1/2}\right)
\label{1s0anal}
\end{equation}
%where
%\begin{equation}
%\lambda = \sqrt{\frac{2\mu V_0 a_m^2}{\hbar^2}}
%\end{equation}
where $\gamma = 0.57721$ is Euler constant and $\epsilon$ is given by:
\begin{equation}
\epsilon = \sqrt{\frac{2\mu E a_m^2}{\hbar^2}} = k a_m
\end{equation}

%\subsubsection{S-wave Energies for \textit{np}-interaction from S-matrix:}
%The low energy SPS are related to scattering parameters through the following effective-range approximation formula \cite{Babenko}:
%\begin{equation}
%    k\cot(\delta)=-\frac{1}{a}+\frac{1}{2}r_ek^2
%    \label{kcot}
%\end{equation}
%where \textit{a} and $r_e$ are scattering length  and effective range respectively.
%Using this approximation, S-matrix can be written in the form \cite{Babenko}:
%\begin{equation}
%    S(k)=\left(\frac{k+i\alpha}{k-i\alpha}\right) \left(\frac{k+i\beta}{k-i\beta}\right)
%    \label{Smatrix}
%\end{equation}
%where 
%\begin{equation}
%    \alpha=\frac{1}{r_e}\bigg[1-\big(1-\frac{2r_e}{a}\big)^{1/2}\bigg]
%    \label{Alpha}
%\end{equation}
%\begin{equation}
%    \beta=\frac{1}{r_e}\bigg[1+\big(1-\frac{2r_e}{a}\big)^{1/2}\bigg]
%    \label{Beta}
%\end{equation}
%Here, $'a'$ is scattering length and $'r_{e}'$ is effective range.
%S-matrix has two poles $i\alpha$ ($\alpha < 0$) and $i\beta$ ($\beta > 0$) , situated in lower and upper half-plane of k respectively, whose energies are given by
%\begin{equation}
%    \epsilon_S=\frac{\hbar^2\alpha^2}{2\mu}
%    \label{Senergy}
%\end{equation}
%\begin{equation}
%    \epsilon_b=\frac{\hbar^2\beta^2}{2\mu}
%    \label{benergy}
%\end{equation}
%
%for the $^{1}S_{0}$ Scattering state and $^{3}S_{1}$ bound state respectively.
%In order to obtain, low energy scattering parameters, SPS at different lab energies upto 10 MeV are utilised. In this paper, we obtain SPS using phase function method.
%

\subsection{Phase Function Method:}
PFM is one of the important tools in scattering studies for both local and non-local interactions \cite{Calogero}. The second order differential equation Eq.\ref{Scheq} can been transformed to a first order non-homogeneous differential equation of Riccati type \cite{Calogero,Babikov}, containing phase shift information, given by: 
%\begin{equation}
%     \frac{d}{dr}tan(\delta_\ell(r))=-\frac{U(r)}{k}\bigg\{\hat{j}_{\ell}(kr)-\hat{\eta}_{\ell}(kr) tan(\eta_\ell(r))\bigg\}^2
%\end{equation}
%where $U(r)=\frac{2\muV(r)}{\hbar^2}$. 
%
%For $\ell = 0$, the Ricatti-Bessel and Riccati-Neumann functions $\hat{j_0}$ and $\hat{\eta_0}$ get simplified as $sin(kr)$ and $-cos(kr)$, so the PF equation for $\ell=0$ becomes
% 
% \begin{equation}
%  \frac{d}{dr}tan(\delta_0(r))=-\frac{U(r)}{k}\bigg\{sin(kr)+cos(kr) tan(\delta_0(r))\bigg\}^2   
% \end{equation}
%  \begin{equation}
%  \frac{1}{cos^2(\delta_0(r))}\frac{d\delta_0(r)}{dr}=-\frac{U(r)}{k}\bigg\{sin(kr)+cos(kr) \bigg(\frac{sin(\delta_0(r))}{cos(\delta_0(r))}\bigg)\bigg\}^2   
% \end{equation}
% or
% \begin{equation}
% \frac{d\delta_0(r)}{dr}=-\frac{U(r)}{k}\bigg\{sin(kr) cos(\delta_0(r))+cos(kr) sin(\delta_0(r))\bigg\}^2      
% \end{equation}
% or
%\begin{equation}
%    \delta_0'(k,r)=-\frac{U(r)}{k}sin^2[kr+\delta_0(r)]
%\end{equation} 
%PFM for $\ell>0$ is  
 
\begin{equation}
\delta_{\ell}'(k,r)=-\frac{U(r)}{k}\bigg[\cos(\delta_\ell(k,r))\hat{j}_{\ell}(kr)-\sin(\delta_\ell(k,r))\hat{\eta}_{\ell}(kr)\bigg]^2
\label{PFMeqn}
\end{equation}
where $U(r)=\frac{2\mu V(r)}{\hbar^2}$.
Prime denotes differentiation of phase shift with respect to distance and Riccati Hankel function of first kind is related to $\hat{j_{\ell}}(kr)$ and $\hat{\eta_{\ell}}(kr)$ by $\hat{h}_{\ell}(r)=-\hat{\eta}_{\ell}(r)+\textit{i}~ \hat{j}_{\ell}(r)$. 
For $\ell = 0$, the Ricatti-Bessel and Riccati-Neumann functions $\hat{j_0}$ and $\hat{\eta_0}$ get simplified as $sin(kr)$ and $-cos(kr)$. So phase equation, for $\ell=0$, is
\begin{equation}
    \delta_0'(k,r)=-\frac{U(r)}{k}sin^2[kr+\delta_0(r)]
\end{equation} 
This is numerically solved using Runge-Kutta 5th order (RK-5) method with initial condition $\delta_{\ell}(k,0)=0$.
%In integral form, the above equation can be written as
%\begin{equation}
%\delta_{\ell}(k,r)=-\frac{1}{k}\int_{0}^{r}{U(r)}\bigg[\cos(\delta_{\ell}(k,r))\hat{j_{\ell}}(kr)-\sin(\delta_{\ell}(k,r))\hat{\eta_{\ell}}(kr)\bigg]^2 dr
%\end{equation}
%The function $\delta_{\ell}(k,r)$ is called the phase function. The  advantage of this method is that, the phase-shifts are directly expressed in terms of the potential and have no relation to the wavefunction. Also, rather than solving the second order Schr$\ddot{o}$dinger equation we only need to solve the first order non-homogeneous differential equation of Riccati type, given by Eq.\ref{PFMeqn}, for phase shift calculations.  
The Ricatti-Bessel and Riccati-Neumann functions can be easily obtained by using following recurrence formulas:
\begin{equation}
    \hat{j_\ell}(kr)=\frac{2\ell+1}{kr} \hat{j_\ell}(kr)-{\hat{j}_{\ell-1}}(kr)
\end{equation}
\begin{equation}
    \hat{\eta_\ell}(kr)=\frac{2\ell+1}{kr} \hat{\eta_\ell}(kr)-{\hat{\eta}_{\ell-1}}(kr)
\end{equation}
%The phase equations for P-wave ($\ell = 1$), D-wave($\ell = 2$), F-wave ($\ell = 3$), G-wave ($\ell = 4$) and H-wave ($\ell = 5$) are given in Appendix-I.

\subsection{Scattering Cross Section:}
Once, SPS $\delta_{\ell}(E)$ are obtained for each orbital angular momentum $\ell$, one can calculate partial SCS $\sigma_{\ell}(E)$ \cite{Amsler}, as :
\begin{equation}
\sigma_{\ell} (E; S,J) = \frac{4\pi}{k^2}\sum_{S=0}^{1}\left(\sum_{J=|\ell - S|}^{|\ell + S|} (2\ell +1)\sin^{2}(\delta_{\ell}(E;S,J))\right)
\label{Pxsec}
\end{equation} 
and total SCS $\sigma_{T}$, is given as
\begin{equation}
\sigma_{T}(E;S,J) =  \frac{1}{\sum_{J=|\ell - S|}^{|\ell + S|}(2J+1)}\sum_{\ell=0}^{5}\sum_{S=0}^{1}(2J+1) \sigma_{\ell}(E;S,J)
\label{Txsec}
\end{equation}

\section{Results and Discussion:}
\begin{figure}[h]
\begin{center}
\includegraphics[scale=0.3]{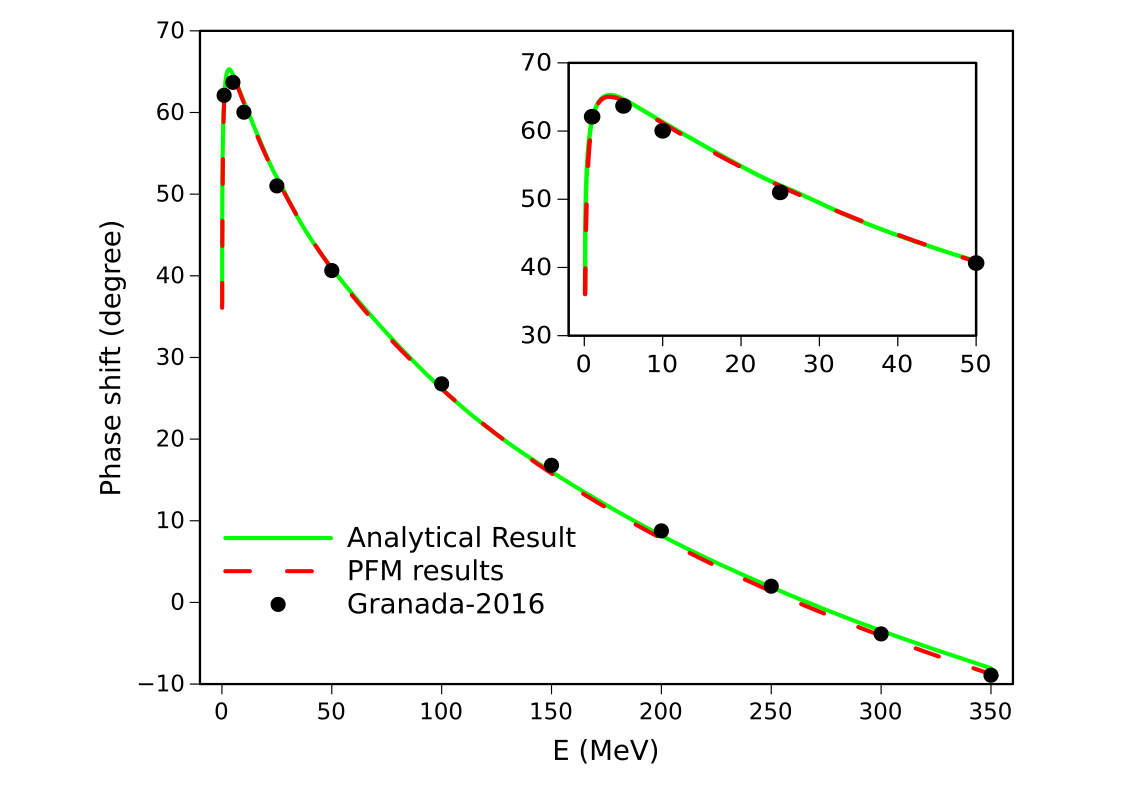}
\caption{Plot of  Scattering phase shifts obtained using analytical and PFM for $^1S_0$-state. The inset shows the match at lower energy upto 50 MeV.}
\label{figure}
\end{center}
\end{figure}

The experimental SPS data have been taken from Perez et.al., of Granada group \cite{Perez}, 2016. To this, we have added an extra data point at lab energy 0.1 MeV for $^{3}S_{1}$ and $^{1}S_{0}$ states, from Arndt (private communication) to improve determination of low energy scattering parameters. The optimised model parameters for all states of different $\ell$-channels are given in Table \ref{table1}.
\begin{table}[h]
\caption{Model Parameters $V_0$ in MeV, $r_m ~\& ~a_m$ in fm for various states of different $\ell$ - channels with obtained mean squared error (MSE) values.}
\scalebox{0.7}{
\setlength{\tabcolsep}{8pt} % Default value: 6pt
\renewcommand{\arraystretch}{1.2}
\begin{tabular}{@{}|rrrrr|rrrrr|rrrrr|@{}} 
\hline
States & $V_0$      & $r_m $    & $a_m $     & $MSE$   & States & $V_0 $      & $r_m$    & $a_m$     & $MSE$    & States & $V_0     $  & $r_m $    & $a_m$    & $MSE$   \\ \hline
$^3S_1$    & 114.153 & 0.841 & 0.350 & 0.155  & $^1D_2$     & 131.302 & 0.010 & 0.526 & 0.026 & $^3F_3$    & 0.010  & 8.807 & 2.441 & 0.025 \\
$^1S_0$    & 70.438  & 0.901 & 0.372 & 0.649 & $^3D_1$    & 0.010   & 7.401 & 1.403 & 0.274 & $^3F_4$    & 40.343 & 0.528 & 0.489 & 0.001 \\
$^1P_1$    & 0.010   & 5.442 & 1.016 & 1.568 & $^3D_2$    & 106.379 & 0.209 & 0.747 & 0.066 & $^1G_4$    & 20.390 & 0.010 & 0.673 & 0.002 \\
$^3P_0$     & 11.579  & 1.750 & 0.601 & 0.049 & $^3D_3$    & 23.620  & 1.185 & 0.305 & 0.002 & $^3G_3$       & 0.010  & 6.846 & 1.486 & 0.009 \\
$^3P_1$     & 0.010   & 4.514 & 0.778 & 0.832 & $^1F_3$    & 0.010   & 7.645 & 1.877 & 0.056 & $^3G_4$       & 59.219 & 0.010 & 0.786 & 0.015 \\
$^3P_2$     & 77.558  & 0.444 & 0.404 & 0.001 & $^3F_2$    & 2.746   & 2.140 & 0.532 & 0.003 & $^3H_4$       & 23.245 & 0.010 & 0.693 & 0.000 \\ \hline
\end{tabular}
}
\label{table1}
\end{table}
Using obtained model parameters, $V_{0} = 70.438,r_{m} = 0.901$ and $a_{m} = 0.372$, for $^{1}S_{0}$ in analytical expression of Eq. \ref{1s0anal}, we obtained corresponding SPS at various energies. These are observed to be closely matching with SPS obtained using PFM as shown in Fig. \ref{figure}, thus validating the later procedure.
The $^3S_1$ model parameters are obtained by utilising energy condition, in Eq \ref{eigenvals} with $v=0$, as a constraint.\\
The obtained SPS for all channels are in good match with experimental data with MSE values being very close to zero. The exceptions are states $^1P_1$, $^3P_1$ and $^3D_1$ with purely negative SPS, where MSE values are 1.568, 0.832 and 0.274 respectively. It is interesting to note that all of them have $V_0$ to be 0.010 MeV and $r_m$ to be large, which makes the shape of their potentials to be of exponential form. Similarly, states $^1F_3$, $^3F_3$ and $^3G_3$, with purely negative SPS, also have $V_0$ to be 0.010 MeV and their MSE slightly higher than other F and G states with positive SPS.
 
%The low energy SPS are related to scattering parameters through the following effective-range approximation formula \cite{Babenko}:
%\begin{equation}
%    k\cot(\delta)=-\frac{1}{a}+\frac{1}{2}r_ek^2
%    \label{kcot}
%\end{equation}
%where \textit{a} and $r_e$ are scattering length  and effective range respectively. 
By using effective-range approximation formula \cite{Babenko} for low energy scattering parameters,  scattering length $'a'$ and effective range $'r_{e}'$ are determined(experimental) to be $-23.390 ~(-23.749(8))~fm$ and $2.42~(2.81(5))~fm$ for $^1S_0$ state and  $5.356~(5.424(3))~fm$ and $1.75~(1.760(5))~fm$ for $^3S_1$ state respectively.        

\begin{figure*}[htp]
\centering
\caption{Inverse Potentials (\textbf{left}) and Scattering phase shifts (\textbf{right}) plots for S, P and D-states of np scattering using RPA.}
\includegraphics[scale=0.32]{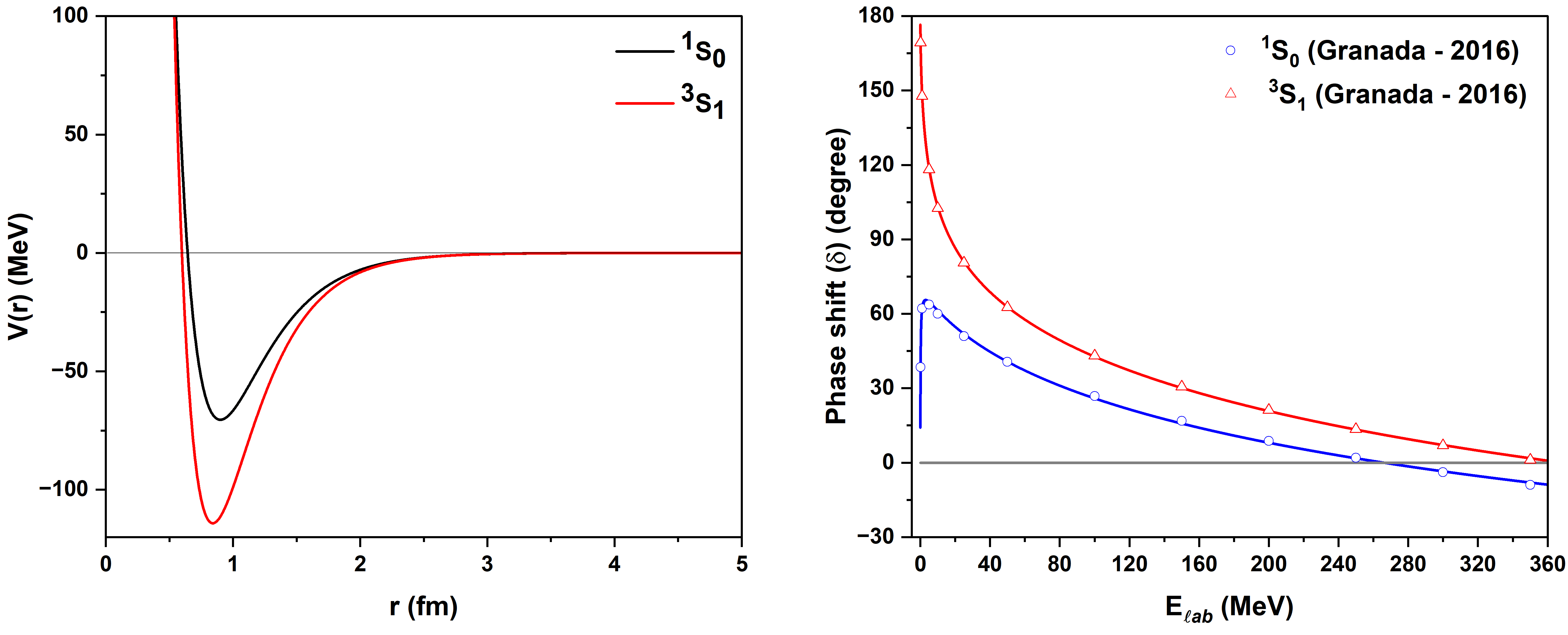}
\includegraphics[scale=0.32]{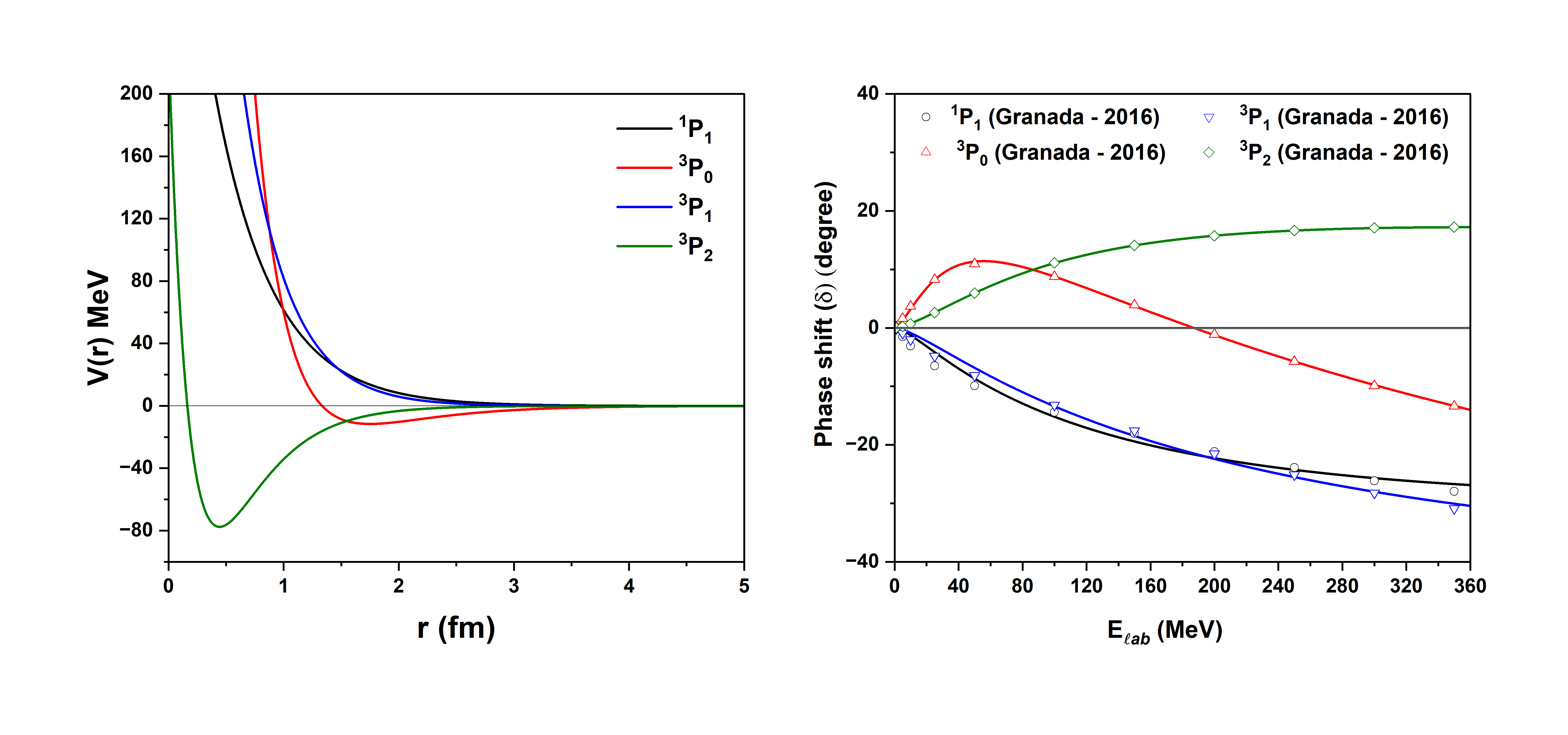}
\includegraphics[scale=0.32]{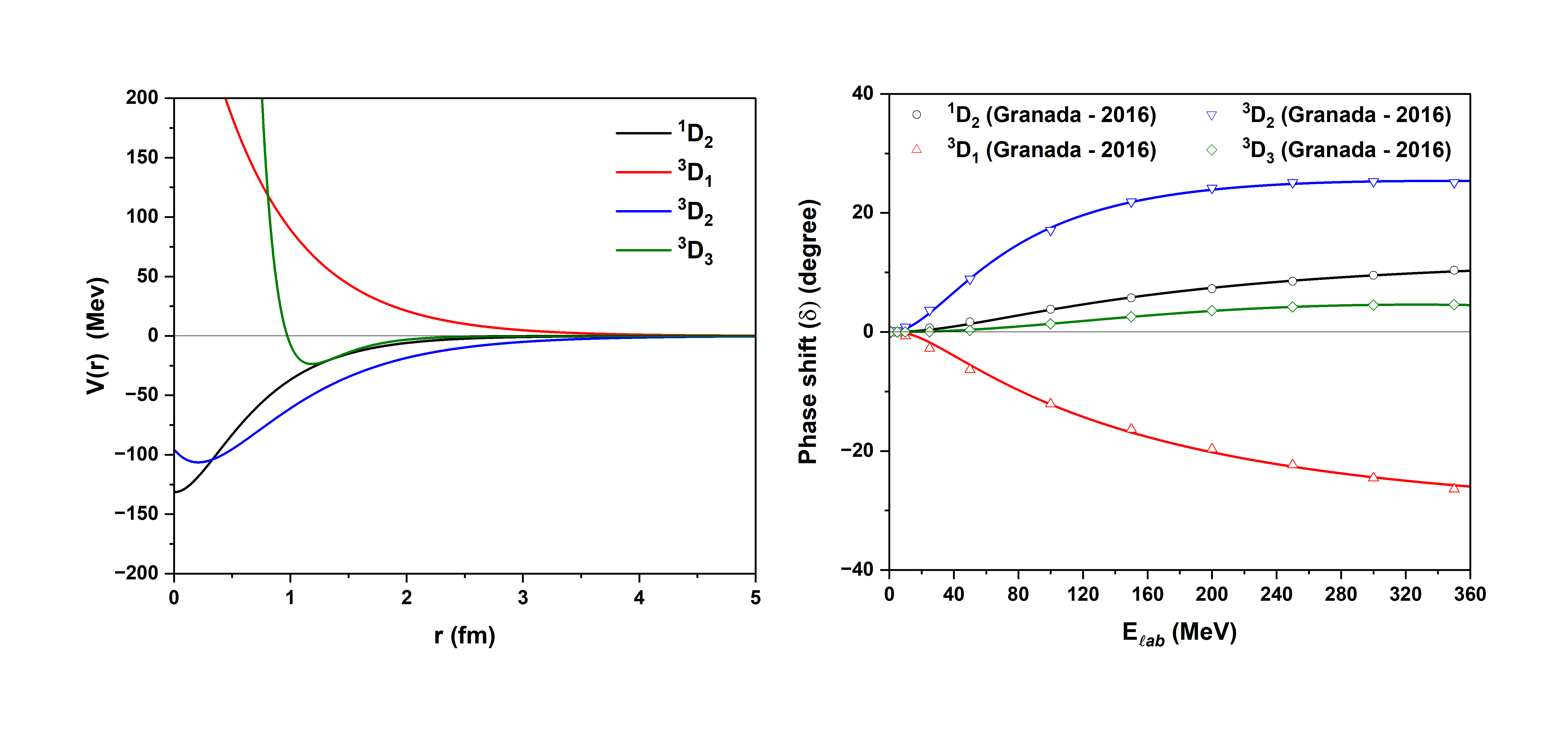}
\label{fig1}
\end{figure*}
\begin{figure*}[htp]
\centering
\caption{Inverse Potentials (\textbf{left}) and Scattering phase shifts (\textbf{right}) plots for F, G and H-states of np scattering using RPA }
\includegraphics[scale=0.32]{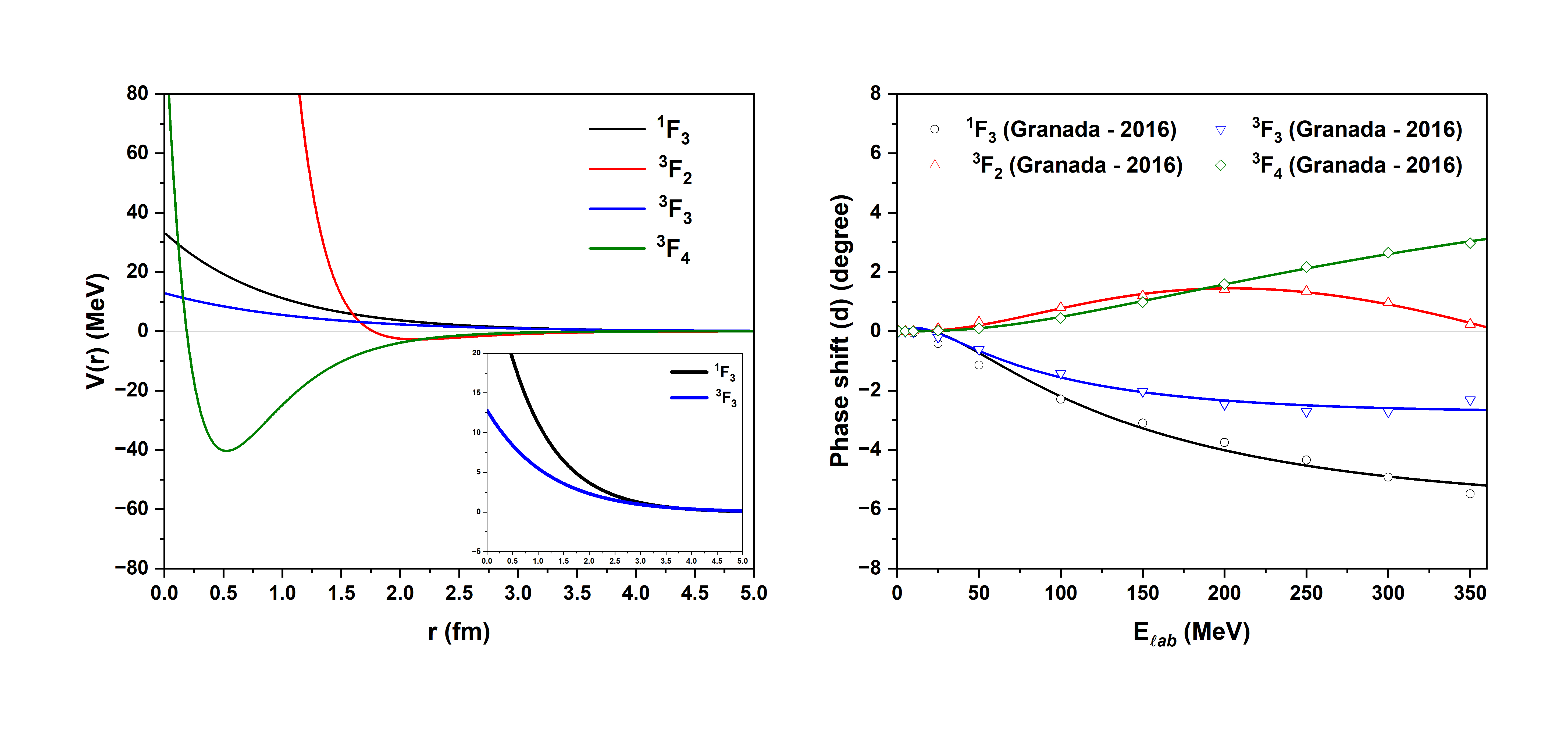}
\includegraphics[scale=0.32]{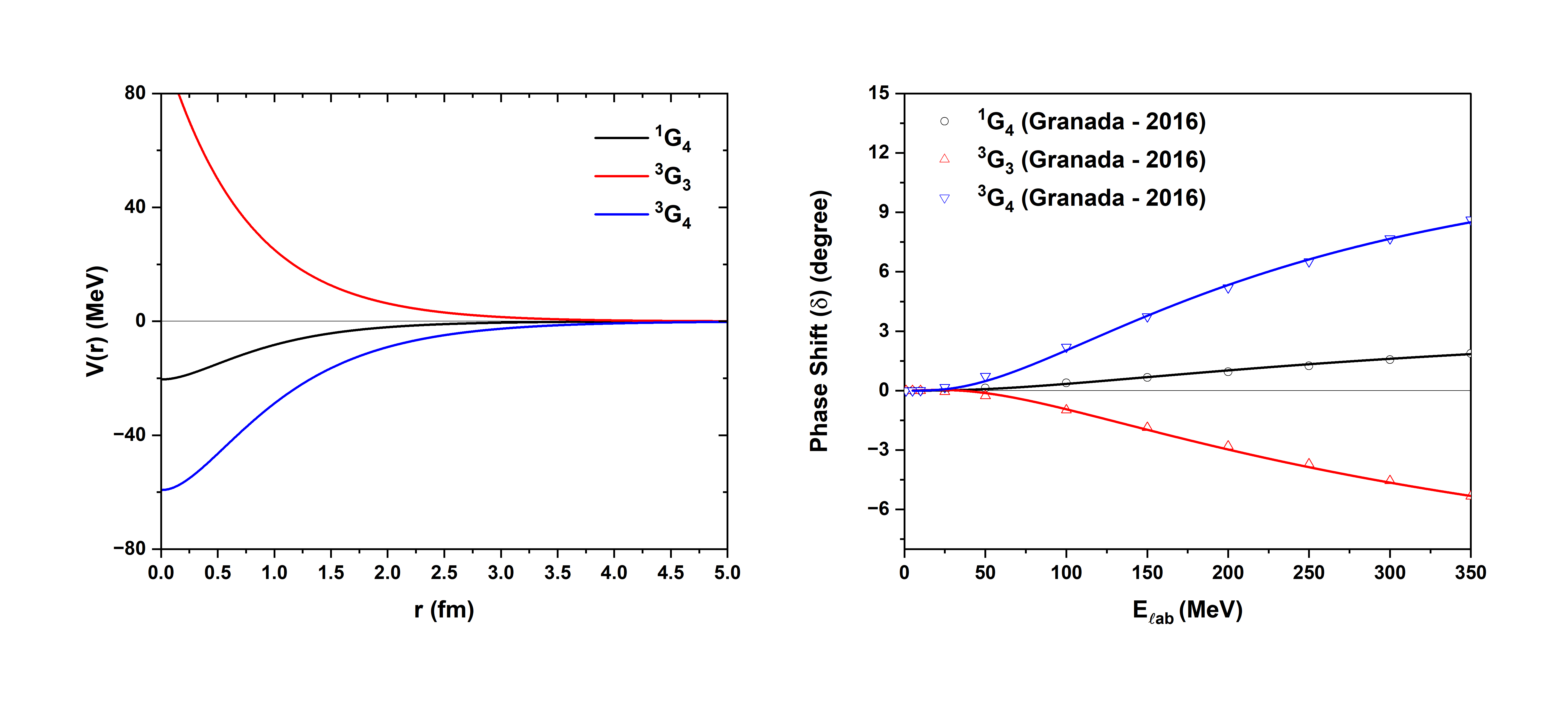}
\includegraphics[scale=0.32]{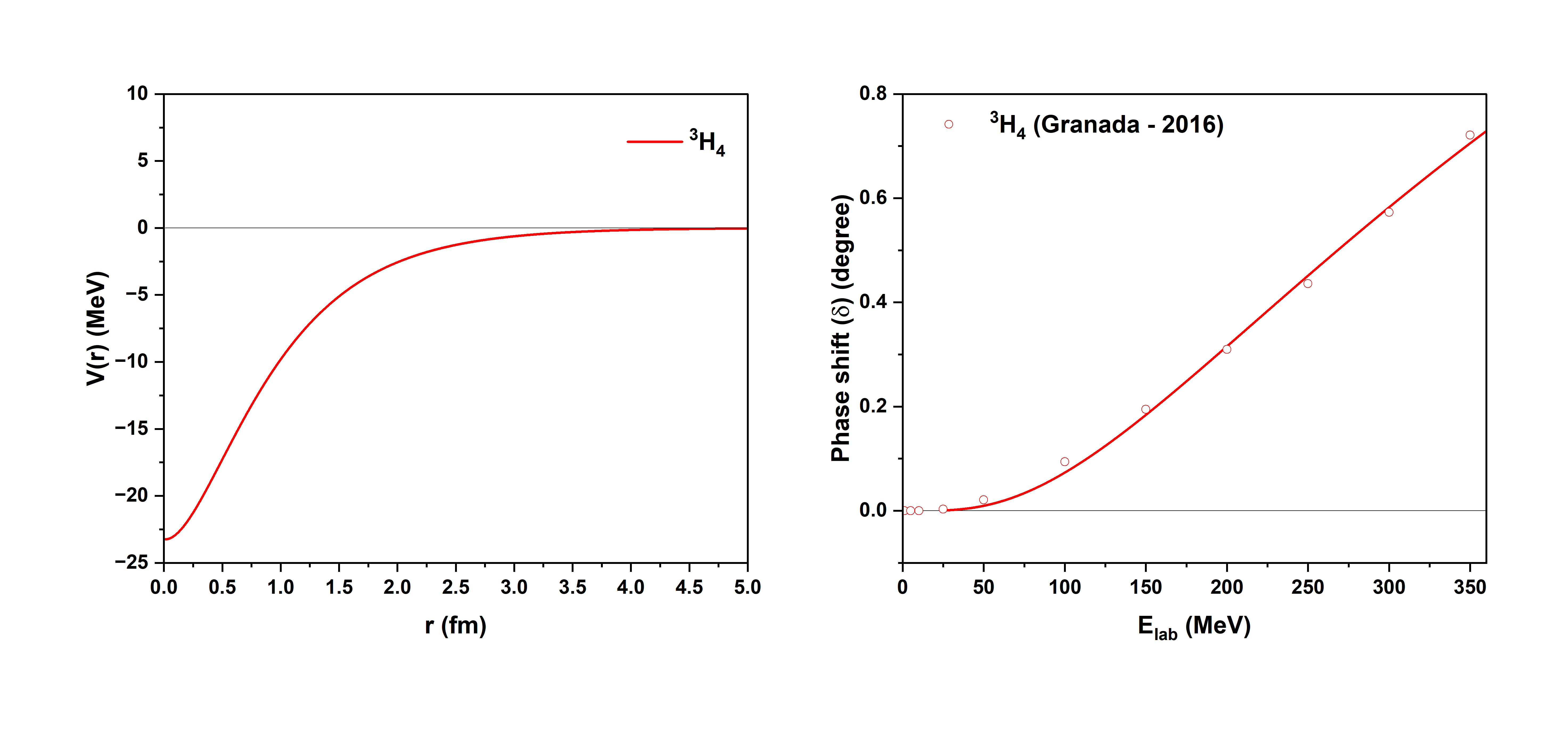}
\label{fig2}
\end{figure*}
\begin{figure}
\includegraphics[scale=0.25]{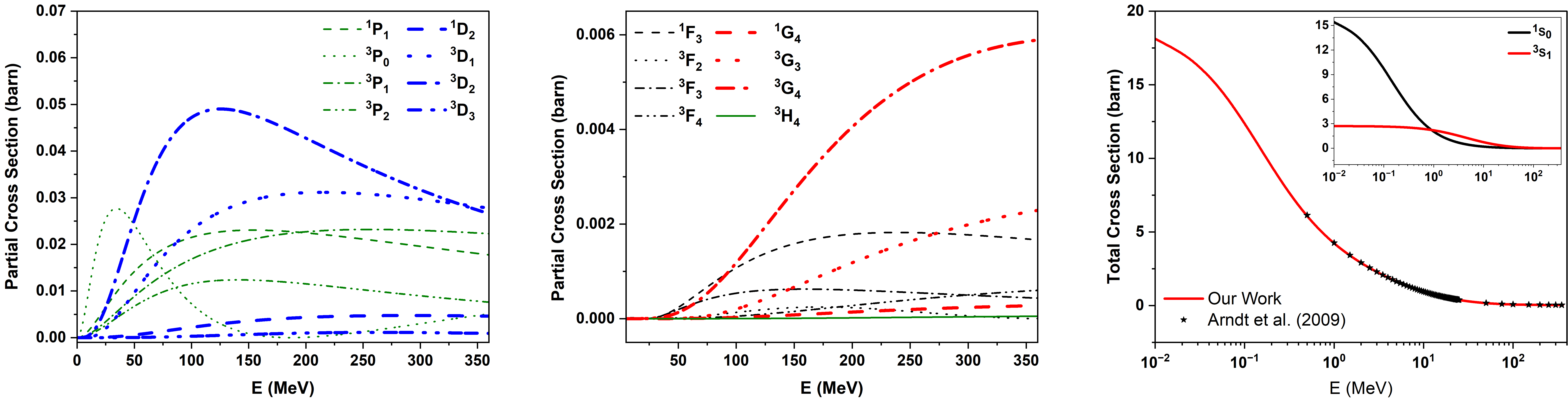}
\caption{Partial scattering cross-section (SCS) due to (a) P and D channels (b) F and G channels, as a function of energy. (c) The total SCS has been plotted on a log scale of energies. Its inset shows contributions due to both S-states.} 
\label{scs}
\end{figure}

The inverse potentials obtained for S, P and D states and their corresponding SPS plots are given in Fig. \ref{fig1}. Similarly, computed inverse potentials and their corresponding SPS plots for F, G and H states of np-interaction are shown in Fig. \ref{fig2}. 
One can observe close match between obtained and expected SPS values for all the states. 
The following observations can be made from these potentials and SPS plots:
\begin{itemize}
\item Both S-state potentials are having similar form with only their depths being different.
\item The potentials have attractive nature whenever SPS are positive and repulsive nature when SPS are negative. 
\item Whenever SPS start from being positive and then cross-over to negative values as seen in $^3P_0$ or dips down as in $^3F_1$ at higher energies, the repulsive nature of the potential curve sets in at higher values of inter-nucleon distance.
\item In case of $^3P_1$ and $^1P_1$, just as their SPS cross-over after 200 MeV, so also their respective inverse potentials cross-over after at about 1.5 fm.
\item It is interesting to note that for certain D-states and all G and H states for which SPS are positive, potential assumes shape of a Gaussian function.
\item All states with negative SPS are having an exponentially decaying positive potential. 
\item Note that, in case of $^1F_3$ and $^3F_3$, the former has more negative SPS than the later and its corresponding inverse potential also has increasing repulsion with decreasing internucleon distances.  
\end{itemize}
The beauty of the method is that Morse function acts as an exponential function for states with negative SPS and a gaussian function for positive ones. When SPS tend to have both negative and postive SPS or those with positive trend to begin with and decrease at higher energies, typical Morse curve is obtained.
Utilising the obtained SPS for various states of all $\ell$-channels, the partial and total scattering cross-sections (SCS) have been calculated using Eqs. \ref{Pxsec} and \ref{Txsec} respectively. In case of $\ell = 0$, the individual partial cross-sections due to both $^1S_0$ and $^3S_1$ have been calculated without summing over their contributions. All these calculated SCS at various energies have been presented in Table \ref{Table 2}. The $\%$-contributions due to individual S-states and rest of the $\ell$-channels from P to H, to the calculated total SCS, are given in brackets. One can observe that $^1S_0$ has large contribution at low energies below 1 MeV and then gradually falls down with increasing energy and becomes very less past 100 MeV. On the other hand, contribution due to $^3S_1$ increases beyond 1 MeV and peaks at 10 MeV and then falls down beyond 250 MeV. The contributions from P and D channels become significant for higher energies from 100 MeV to 350 MeV. Those due to F and G are far less in comparision in the same range but certainly important for accurately describing the observed experimental total SCS. The SPS are available for only 1 H-state and it's contribution is almost negligible to the determination of total SCS. Overall the obtained total SCS are found to be closely matching the experimental ones. The partial cross-sections due to P and D channels are shown in Fig \ref{scs}(a) and those due to F and G channels in Fig \ref{scs}(b) with respect to lab energies. The total SCS has been plotted with respect to lab energies on a log scale in Fig. \ref{scs}(c) with individual contributions due to individual S-states as an inset. 

\begin{table}[h]
\caption{Individual contributions to calculated total elastic scattering cross-section (SCS) from various channels. In case of $\ell=0$, the contributions due to both $^1S_0$ and $^3S_1$ are given seperately. The $\%$-contributions to obtained total SCS has been given in brackets.}
\scalebox{0.7}{
\setlength{\tabcolsep}{7pt} % Default value: 6pt
\renewcommand{\arraystretch}{1.2}
\begin{tabular}{@{}llllllllll@{}}
\toprule
E & $\sigma_{exp}$(b)  & $\sigma_{^{1}S_{0}}$  &$\sigma_{^{3}S_{1}}$ & $\sigma_{P}$ & $\sigma_{D}$ & $\sigma_{F}$&  $\sigma_{G}$  &  $\sigma_{H}$&$\sigma_{sim}$(b) \\ 
(MeV) & \cite{Arndt} & & & & & \\ \midrule
0.1         & -              & 9.708(78\%)   & 2.651 (22\%)   & 2.31$\times 10^{-7}$ & 2.98$\times 10^{-12}$ & 0        & 0        & 0        & 12.359         \\
0.5         & 6.135          & 3.653 (60\%)   & 2.425 (40\%)   & 5.690$\times 10^{-6}$ & 1.78$\times 10^{-9}$ & 0        & 0        & 0        & 6.078           \\
1           & 4.253          & 2.041 (48\%)   & 2.189 (52\%)    & 2.240$\times 10^{-5}$ & 2.69$\times 10^{-8}$ & 0        & 0        & 0        & 4.230          \\
10          & 0.9455         & 0.2007 (21\%)  & 0.7413 (79\%)  & 0.0016   & 0.0001   & 6.40$\times 10^{-6}$ & 4.81$\times 10^{-8}$ & 0        & 0.9437         \\
50          & 0.1684         & 0.0222(13\%)  & 0.1235 (75\%)  & 0.0106 (6\%)   & 0.0079 (5\%)  & 0.0001   & 3.44$\times 10^{-5}$ & 1.32$\times 10^{-8}$ & 0.1643         \\
100         & 0.07553        & 0.00498 (7\%) & 0.03601(50\%) & 0.015(21\%)    & 0.01593 (22\%)  & 0.00044 (1\%) & 0.00034  & 3.84$\times 10^{-7}$ & 0.07270        \\
150         & 0.05224        & 0.00129(3\%) & 0.01314 (26\%) & 0.01622 (33\%) & 0.01752 (35\%)  & 0.00064 (1\%) & 0.00083 (2\%)  & 1.61$\times 10^{-6}$ & 0.04964        \\
200         & 0.04304        & 0.00025(1\%) & 0.00493 (12\%) & 0.0161 (40\%)   & 0.01679 (42\%)  & 0.00073 (2\%)  & 0.00128 (3\%)  & 3.56$\times 10^{-6}$ & 0.04008        \\
250         & 0.03835        & 0.00001 & 0.00166 (5\%) & 0.01546 (44\%) & 0.01542 (44\%) & 0.00075 (2\%) & 0.00162 (5\%)  & 5.82$\times 10^{-6}$ & 0.03493        \\
300         & 0.03561        & 0.00003 & 0.00040(1\%) & 0.01464(46\%)  & 0.01394(44\%)  & 0.00074 (2\%)  & 0.00184 (6\%) & 8.09$\times 10^{-6}$ & 0.03160        \\
350         & 0.03411        & 0.00015 & 0.00002 & 0.01377 (47\%) & 0.01255(43\%)  & 0.00072 (2\%)  & 0.00197(7\%)  & 0.00001  & 0.02919        \\ \bottomrule
\label{Table 2}
\end{tabular}
}

\end{table}

\section{Conclusions:}
The inverse potentials for np-interaction for all partial waves with $\ell$ = 0 to 5 have been constructed using reference potential approach, by choosing Morse function as zeroth reference, for the first time. The model parameters have been obtained by choosing to minimize mean squared error between SPS obtained from PFM technique and Granada data \cite{Perez} in an iterative manner. This was achieved by making suitable adjustments to different model parameters using optimisation routines. The obtained SPS for all the channels match expected ones very closely. Overall, reference potential approach using Morse function seems to be able to give reasonably accurate inverse potentials of appropriate shapes that could logically explain observed trends in SPS for variour $\ell$-channels, as expected from PFM. The total scattering cross-sections have been determined for data upto 350 MeV and are found to be in good agreement with experimental values. It remains to be seen as to how to construct inverse potentials for pp-scattering using RPA. 

\section*{Acknowledgments}
A. Awasthi acknowledges financial support provided by Department of Science and Technology
(DST), Government of India vide Grant No. DST/INSPIRE Fellowship/2020/IF200538. The corresponding author dedicates this work to his inspirational guide Padma Shri Prof P. C. Sood with gratitude. \\
 The authors declare that they have no conflict of interest. 

%\section*{Appendix-I}
% PF equation for $\ell=1, 2, 3~\&~4$ are obtained as follows:
%\begin{equation}
%\delta_1'(k,r)=-\frac{U(r)}{k}\bigg[\frac{sin(\delta_1+\kappa)-\kappa cos(\delta_1+\kappa)}{\kappa}\bigg]^2
%\end{equation}
%
%
%\begin{equation}
%\delta_2'(k,r) = -\frac{U(r)}{k}\bigg[-\sin{\left(\delta_2+ \kappa \right)}-\frac{3 \cos{\left(\delta_2 + \kappa \right)}}{\kappa} + \frac{3 \sin{\left(\delta_2 + \kappa \right)}}{\kappa^2}\bigg]^2 
%\end{equation} 
%
%\begin{eqnarray}
%\delta_3'(k,r)=-\frac{U(r)}{k}\bigg[{cos(\delta_3+ \kappa)-\frac{6}{\kappa}sin(\delta_3+ \kappa)-\frac{15}{\kappa^2} cos(\delta_3+ \kappa)+\frac{15}{\kappa^3}sin(\delta_3+kr)}\bigg]^2   
%\end{eqnarray}
%
%\begin{eqnarray}\nonumber
%\delta_4'(k,r)=-\frac{U(r)}{k}\bigg[\sin{\left(\delta_4 + \kappa \right)} + \frac{10 \cos{\left(\delta_4 + \kappa \right)}}{\kappa}-\frac{45 \sin{\left(\delta_4 + \kappa \right)}}{ \kappa^2}\\~~~~~~~~~~~~~- \frac{105 \cos{\left(\delta_4 + \kappa \right)}}{ \kappa^3}+ \frac{105 \sin{\left(\delta_4 + \kappa \right)}}{ \kappa^4}\bigg]^2
%\end{eqnarray}
%
%
%\begin{eqnarray}\nonumber
%\delta_5^{\prime}(k, r)= -\frac{U(r)}{k}\left[\left(21\left(\kappa^4-60 \kappa^2+495\right) \kappa \cos (\kappa)+\left(\kappa^6-210 \kappa^4+4725 \kappa^2-10395\right)\sin (\kappa)\right)\right.\\\nonumber
%\cos (\delta_5)-\left(21\left(\kappa^4-60 \kappa^2+495\right) \kappa \sin (\kappa)-\left(\kappa^6-210 \kappa^4+4725 \kappa^2-10395\right) \cos (\kappa)\right)
%\sin (\delta_5)]^2\\ \times \frac{1}{\kappa^{10}}
%\end{eqnarray}

\end{document}